\documentclass[aip,apl,twocolumn,preprintnumbers,amsmath,amssymb]{revtex4}

\usepackage{graphicx}
\usepackage[colorlinks=true,citecolor=blue]{hyperref}

\begin{document}
    \title{Phonon-mediated coupling between quantum dots through an off-resonant microcavity}
    \author{Arka Majumdar}
    \email{arkam@stanford.edu}
    \author{Michal Bajcsy}
    \author{Armand Rundquist}
    \author{Erik Kim}
    \author{Jelena Vu\v{c}kovi\'{c}}
    \affiliation{E.L.Ginzton Laboratory, Stanford University, Stanford, CA, $94305$\\}
\begin{abstract}
We present experimental results showing phonon-mediated coupling
between two quantum dots embedded inside a photonic crystal
microcavity. With only one of the dots being spectrally close to
the cavity, we observe both frequency up-conversion and
down-conversion of the pump light via a $\sim1.2$ THz phonon. We
demonstrate this process for both weak and strong regimes of
dot-cavity coupling, and provide a simple theoretical model
explaining our observations.
\end{abstract}
\maketitle
\section{Introduction}
Phonon-mediated coupling between a self-assembled semiconductor
quantum dot (QD) and a semiconductor microcavity is a recently
discovered phenomenon unique to solid state cavity quantum
electrodynamics \cite{article:michler09,article:majumdar09}. Due
to the electron-phonon interaction in the QD, we observe cavity
emission when the QD is resonantly excited. Alternatively, we see
emission from the QD when the cavity is pumped on resonance. Apart
from the fundamental interest in identifying the mechanism behind
this unusual off-resonant coupling
\cite{kapon_NRDC,article:immamoglu09}, this effect can be used to
probe the coherent interaction of the QD with a strong laser
\cite{majumdar_QD_splitting} and the cavity-enhanced AC stark
shift of a QD \cite{edo_Stark}. These results demonstrate that the
off-resonant cavity constitutes an efficient read-out channel for
the QD states. 

However, all experiments reported so far in the
literature are based on the interaction between a single QD and a
cavity. Recently, an experimental study of two spatially separated QDs interacting
resonantly in a microcavity has been reported
\cite{laucht_two_QD}, as well as a theoretical analysis
\cite{two_qd_theory} of the possible energy transfer mechanisms
between QDs in such a cavity. In this work, we show that two
spectrally detuned QDs can interact with each other via an
off-resonant cavity. More specifically, we observe emission from a
spectrally detuned QD when another QD is resonantly excited. Both
frequency down-conversion (transfer of photons from a higher
energy QD to a lower energy QD) and up-conversion (transfer of
photons from a lower energy QD to a higher energy QD) are
observed for frequency separation of up to $\sim\pm 1.2$ THz. Such a large energy difference cannot be ascribed to an excited state of the same QD, as opposed to earlier work by Flagg et al \cite{shih_phonon} which was performed without a cavity and for frequency difference of $\sim\pm 0.2$ THz. Based on our observations, we believe this process
is happening between two different QDs, and the coupling between
the QDs is enhanced by the presence of the cavity.

\section{Theory}
\begin{figure*}
\centering
\includegraphics[width=6.5in]{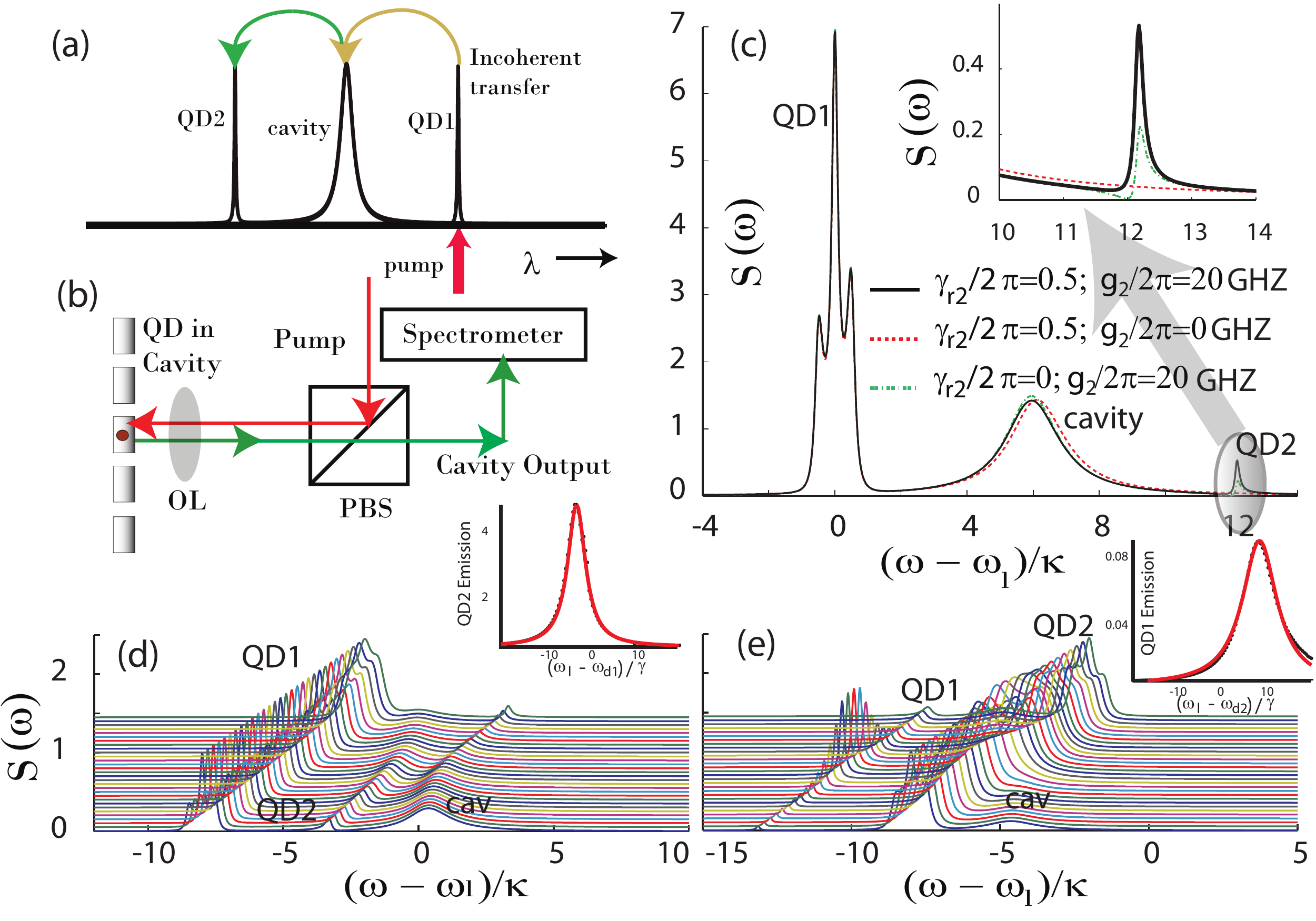}
\caption{(color online) (a) Illustration of the dot to dot coupling.
QD1 is resonantly driven with a pump laser. The excitation is
transferred to QD2 via the cavity. (b) Experimental
cross-polarized confocal microscopy setup. PBS: polarizing beam
splitter, OL: objective lens. (c) Numerically calculated power spectral density
$S(\omega)$ of the QD resonance fluorescence collected through the
cavity for different $\gamma_{r2}$ and $g_2$, when QD1 is
resonantly excited (i.e., $\omega_l=\omega_{d1}$). Inset shows a zoom-in of the QD2 emission. (d), (e)
Calculated $S(\omega)$ when QD1(d) and QD2(e) are resonantly
excited, and the dot-cavity detunings are changed. In other words, $\omega_l=\omega_{d1}$ for (d) and $\omega_l=\omega_{d2}$ for (e). Both QD1 and
QD2 resonances are shifted. We observe anti-crossing between QD2
and the cavity in the off-resonant emission (d). We also observe
an increase in QD1 emission when QD2 is resonant with the cavity
(e). For (d) and (e) parameters used for the simulations were:
$\gamma_1/2\pi=\gamma_2/2\pi=1$GHz,
$\gamma_{r1}/2\pi=\gamma_{r2}/2\pi=0.5$ GHz,
$g_1/2\pi=g_2/2\pi=20$ GHz, $\kappa/2\pi=20$ GHz, detuning between
two dots $5\kappa$, driving laser strength $\Omega_0/2\pi=5$ GHz.}
\label{Fig_th_schematic}
\end{figure*}

The experimental system we want to model is shown in Fig.
\ref{Fig_th_schematic}a. QD1, spectrally detuned from both the
cavity and QD2, is resonantly excited with a pump laser. The
excitation is transferred to the cavity and QD2 via an incoherent
phonon-mediated coupling \cite{majumdar_phonon_11}. We note that,
in theory, it is possible to transfer energy directly from QD1 to
QD2 via phonons. However, we observe the QD2 emission to be
strongly dependent on the QD2-cavity detuning, and hence the
presence of a cavity is important for our experiment. In
particular, for detunings greater than a few cavity linewidths,
the QD2 emission becomes weak and eventually vanishes.

The Master equation to describe the lossy dynamics of the density
matrix $\rho$ of a coupled system consisting of two QDs and a
cavity is given by
\begin{equation}
\frac{d\rho}{dt}=-i[\mathcal{H},\rho]+2\kappa\mathcal{L}[a]+2\gamma_1\mathcal{L}[\sigma_1]+2\gamma_2\mathcal{L}[\sigma_2].
\end{equation}
Assuming the rotating wave approximation, the Hamiltonian describing the coherent
dynamics of the system $\mathcal{H}$ can be written in the interaction picture as
\begin{eqnarray}
\mathcal{H}=\omega_c a^\dag
a&+&\omega_{d1}\sigma_1^\dag\sigma_1+g_1(a^\dag\sigma_1+a\sigma_1^\dag)\\
\nonumber
&+&\omega_{d2}\sigma_2^\dag\sigma_2+g_2(a^\dag\sigma_2+a\sigma_2^\dag),
\end{eqnarray}
while the Lindblad operator modeling the incoherent decay via a
collapse operator $D$ is $\mathcal{L}[D]=D\rho
D^\dag-\frac{1}{2}D^\dag D \rho-\frac{1}{2}\rho D^\dag D$.
Additionally, $\kappa$ is the cavity field decay rate; $\gamma_1$
and $\gamma_2$ are the QD dipole decay rates; $\omega_c$,
$\omega_{d1}$ and $\omega_{d2}$ are the resonance frequencies of
the cavity, QD1 and QD2; $g_1$ and $g_2$ are the coherent
interaction strengths between the cavity and the two QDs. The resonant driving of QD1 or QD2 can be described, respectively, by adding the term
$\Omega(\sigma_1+\sigma_1^\dag)$ or $\Omega(\sigma_2+\sigma_2^\dag)$ to the Hamiltonian $\mathcal{H}$.
The driving laser frequency is denoted by $\omega_l$. We model the incoherent phonon-mediated coupling by adding
$2\gamma_{r1}\mathcal{L}[a^\dag\sigma_1]$ and
$2\gamma_{r2}\mathcal{L}[a\sigma_2^\dag]$ to the Master equation.
The channel between QD1 and the cavity is then characterized by
the values of $\gamma_{r1}$ and $g_1$, while the channel between
the cavity and QD2 is characterized by $\gamma_{r2}$ and $g_2$.
Fig. \ref{Fig_th_schematic}c shows the numerically simulated power
spectral density (PSD) of the QD resonance fluorescence
$S(\omega)=\int_{-\infty}^{\infty}\langle a^\dag(\tau)a(0)\rangle
e^{-i\omega\tau}d\tau$ collected through the cavity. We use only
the cavity operator to calculate the PSD because experimentally
most of the collected light is in the cavity mode. For these
simulations, we use $\gamma_1/2\pi=\gamma_2/2\pi=1$GHz,
$\gamma_{r1}/2\pi=0.5$ GHz, $g_1/2\pi=20$ GHz, $\kappa/2\pi=20$
GHz, QD1-cavity detuning $\Delta_1=6\kappa$ and QD2-cavity
detuning $\Delta_2=-6\kappa$ and the driving laser strength
$\Omega_0/2\pi=5$ GHz.

We first study the role of $\gamma_{r2}$ and $g_2$ in the QD2
emission. Without $g_2$, no emission from QD2 is observed; in the
presence of $g_2$, QD2 emission appears and $\gamma_{r2}$ enhances
it (Fig. \ref{Fig_th_schematic} c). This shows that
coherent coupling between the cavity and QD2 is required to
observe this dot to dot coupling. The three peaks observed at the
QD1 resonance are the usual Mollow triplet, modified due to the
presence of the cavity and phonons \cite{hughes_mollow,
michler_mollow,majumdar_phonon_11}.

Next, we analyze the dependence of the inter-dot coupling on the
spectral detuning between the undriven dot and the cavity. In an
actual experiment  it is very difficult to tune only one QD
without affecting the other, as the two QDs are spatially very
close to each other. Hence, in the simulation, we changed both QD
resonances and kept the cavity resonance fixed. In Fig.
\ref{Fig_th_schematic}d we excite  QD1, which is spectrally far
detuned from the cavity. QD2 is spectrally close to the cavity,
and strongly coupled. The resonant excitation of QD1 causes light
to be emitted both from the cavity and from QD2. Additionally, we
observe anti-crossing between the cavity and QD2 as the frequency
of QD2 is tuned. Following this, we excite QD2 resonantly, and
observe emission from QD1 (Fig. \ref{Fig_th_schematic}e). We
observe an increase in QD1 emission intensity when QD2 is resonant
with the cavity. Finally, we calculate the linewidth of QD1 while
measuring the emission from QD2 (inset of Fig.
\ref{Fig_th_schematic}d), as well as the linewidth of QD2 while
measuring the emission from QD1 (inset of Fig.
\ref{Fig_th_schematic}e), for a weak excitation laser power
($\Omega_0/2\pi=1$ GHz). We find that the linewidth of QD1 is $~4$
GHz and the linewidth of QD2 is $~9$ GHz. These simulated
linewidths are larger than the linewidths one would expect based
on the decay rates, i.e.  $2(\gamma+\gamma_r)/2\pi=3$ GHz, and are
the result of the cavity (with a linewidth of $~40$ GHz) assisting
in the coupling between QD1 and QD2.

\section{Temporal Dynamics of Dot-Cavity Off-Resonant Coupling}
In this section, we describe an experiment to estimate the time
required to transfer the energy from a QD to the cavity, when the
QD is resonantly excited. This measurement gives a way to estimate
the incoherent coupling rate $\gamma_r$. The experiments are
performed in a helium-flow cryostat at cryogenic temperatures
($\sim 30-55$ K) on self-assembled InAs QDs embedded in a GaAs
three-hole defect $L_3$ photonic crystal cavity
\cite{article:eng07}. The $160$nm GaAs membrane used to fabricate
the photonic crystal is grown by molecular beam epitaxy on top of
a GaAs $(100)$ wafer. The GaAs membrane sits on a $918$ nm
sacrificial layer of Al$_{0.8}$Ga$_{0.2}$As. Under the sacrificial
layer, a $10$-period distributed Bragg reflector consisting of a
quarter-wave AlAs/GaAs stack is used to increase the signal
collection into the objective lens. The photonic crystal was
fabricated using electron beam lithography, dry plasma etching,
and wet etching of the sacrificial layer \cite{article:eng07}.

\begin{figure}
\centering
\includegraphics[width=3.5in]{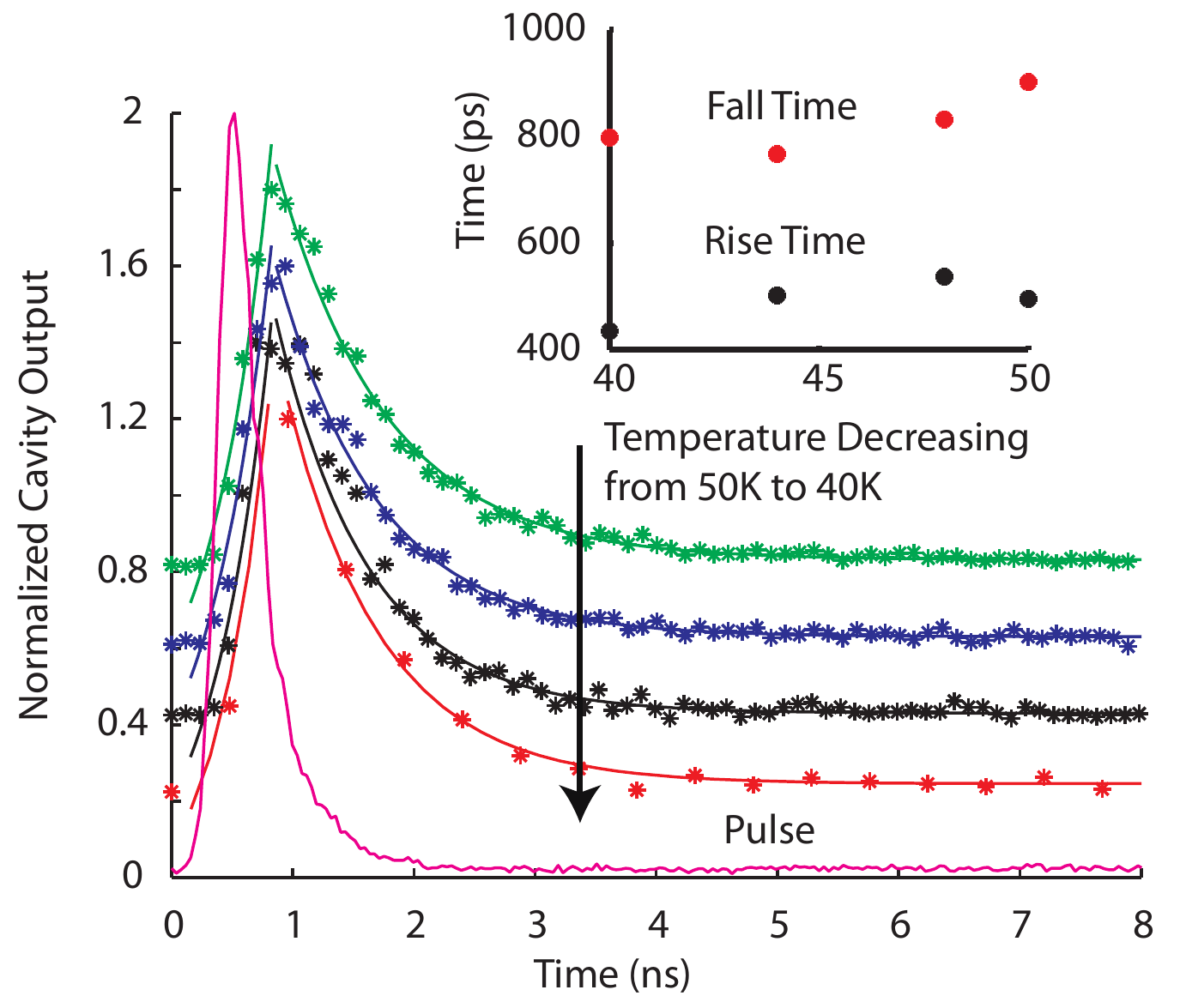}
\caption{(color online) Time resolved measurement of the
off-resonant cavity emission. The QD is resonantly excited with a
$40$ ps pulse, and the time-resolved measurement of the off-resonant cavity emission is performed. The inset plots the rise times and fall times of the
cavity emission (extracted from the exponential fits) against the
system's temperature.} \label{Fig_exp_temporal}
\end{figure}

We resonantly excite the QD with a laser pulse train consisting of
$\sim 40$ ps wide pulses with a repetition period of $13$ ns. A
grating filter is used to collect only the off-resonant cavity
emission and block all the background light from the excitation
laser. The cavity emission signal is then sent to a single photon
counter followed by a picosecond time analyzer (PTA). The PTA is triggered by the excitation
laser pulse, and the cavity emission is recorded. Fig.
\ref{Fig_exp_temporal} shows the pulse shape, as well as the
rising and falling edges of the cavity emission for different
temperatures. By fitting exponentials to the cavity signal, we
estimate the rise and fall times of the cavity emission. The rise
time gives an estimate of the excitation transfer time $\tau_r$
between the QD and the cavity, and in our system it is on the
order of $\sim 500$ ps. From this we use the formula
$\gamma_r\bar{n}=1/\tau_r$ to roughly estimate $\gamma_r/2\pi$ to
be $0.25$ GHz, assuming the mean phonon number $\bar{n}\approx8$. The QD in this
particular case is blue detuned from the cavity, so to have
off-resonant coupling, a phonon needs to be absorbed. The
temperature is changed from $40$ K to $50$ K, corresponding to a
change in dot-cavity detuning from $1.8$ nm to $2.25$ nm and a
change in mean phonon number $\bar{n}$ from $7.8$ to $8$. Similar
values of rise time are obtained for several different QDs, both
blue and red detuned from the cavity. The fall
time corresponds to the cavity life-time, but the measured values
of $\sim 850$ ps are rather large compared to the values
previously reported \cite{arka_switching}. This might be due to
the high density of QDs in the sample, as well as the presence of
phonons.

\section{Coupling between two quantum dots}
\begin{figure}
\centering
\includegraphics[width=3.5in]{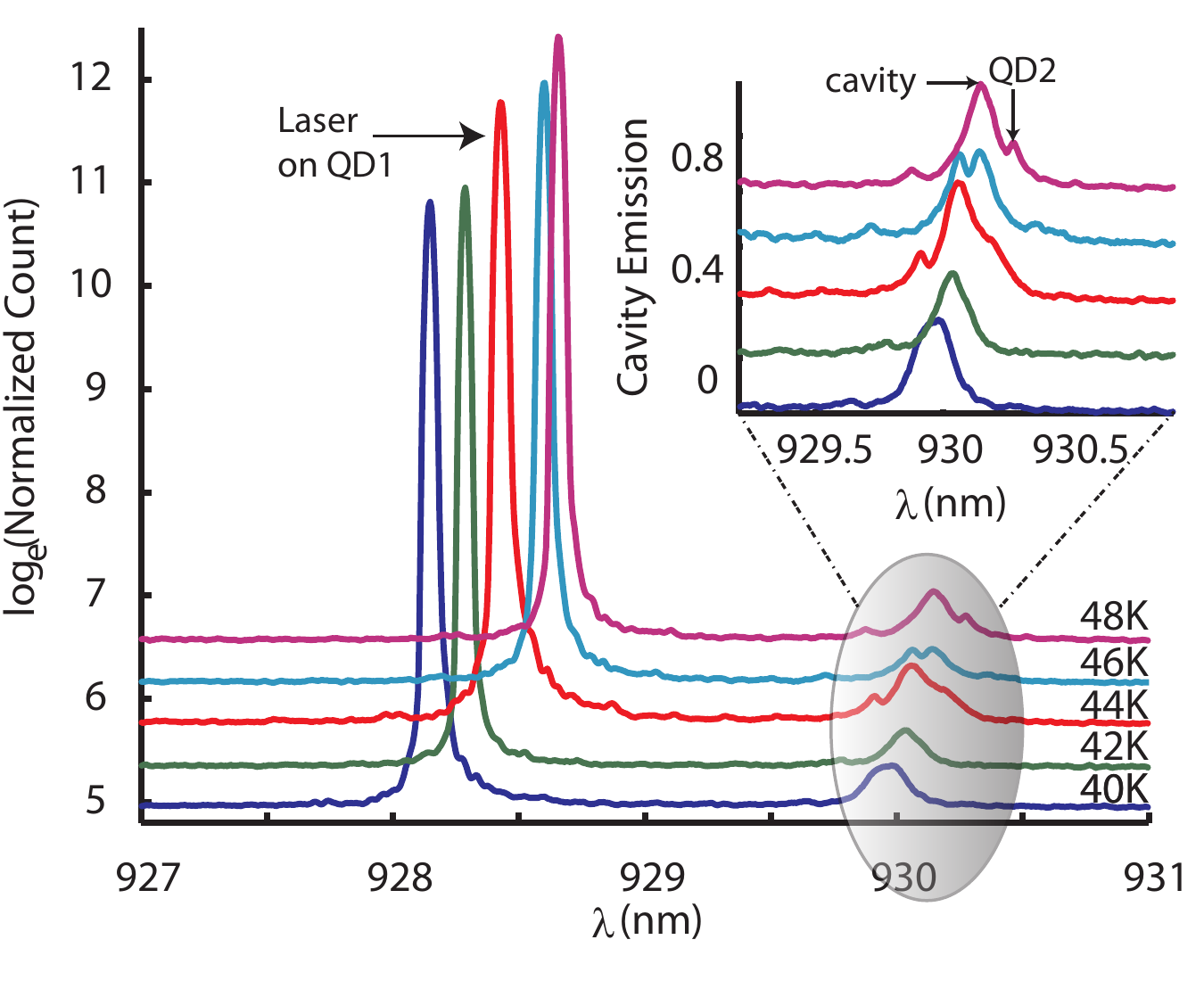}
\caption{(color online) Measurement of the emission from the off-resonant cavity and
QD2, under resonant excitation of QD1. We observe anti-crossing
between QD2 and the cavity when the temperature of the system is
changed.} \label{Fig_exp_sc_dot_temp}
\end{figure}

In this section, we present experimental data showing dot to dot
coupling via an off-resonant cavity for two different systems: one
with a strongly coupled QD, and the other with a weakly coupled
QD. In the first system, we excite QD1 resonantly with a laser,
and observe emission both from the off-resonant cavity and QD2
(Fig. \ref{Fig_exp_sc_dot_temp}). Note that QD2 is strongly
coupled to the cavity and we observe anti-crossing between the
cavity and QD2 in the off-resonant emission when the temperature
of the system is changed (inset of Fig.
\ref{Fig_exp_sc_dot_temp}). The experimental data match well
qualitatively with the theoretical result shown in Fig.
\ref{Fig_th_schematic}d. The emission from QD2 diminishes as QD2
is detuned from the cavity,  which shows that the coupling between
the two dots is enhanced by the presence of the cavity. However,
when we scan the pump laser across QD2 and observe QD1 emission in
this system, we obtain the cavity linewidth showing the usual
cavity to QD1 coupling \cite{article:majumdar10}. This might be
due to the high temperature ($40$-$48$K) of the system, as will be
explained later in this paper.

\begin{figure}
\centering
\includegraphics[width=3.5in]{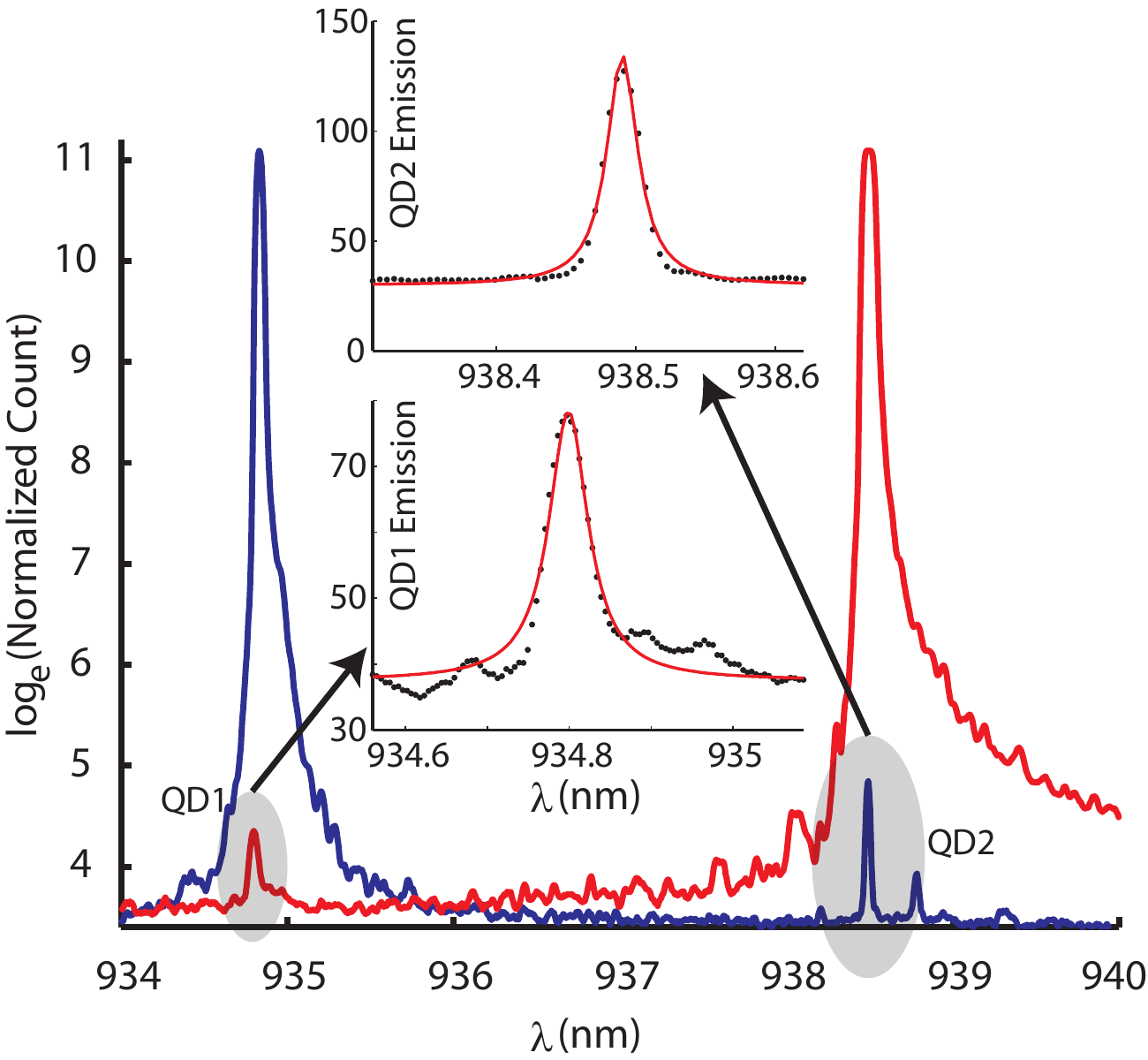}
\caption{(color online) Experimental demonstration of the phonon mediated inter-dot coupling: we
observe the emission from a lower energy QD, when a higher energy
quantum dot is resonantly excited (blue). Similarly, under resonant
excitation of a lower energy QD, emission from a higher energy QD
is observed (red). Inset zooms into the QD emission. QD linewidths are
estimated by fitting Lorentzians. Measured linewidths of the higher and
lower energy QDs, respectively, are $\sim 0.03$ nm and $\sim
0.013$ nm. The cavity is at $\sim 935$ nm, close to the higher
energy QD. } \label{Fig_exp_dot_dot_coupling_part_a}
\end{figure}

\begin{figure}
\centering
\includegraphics[width=3.5in]{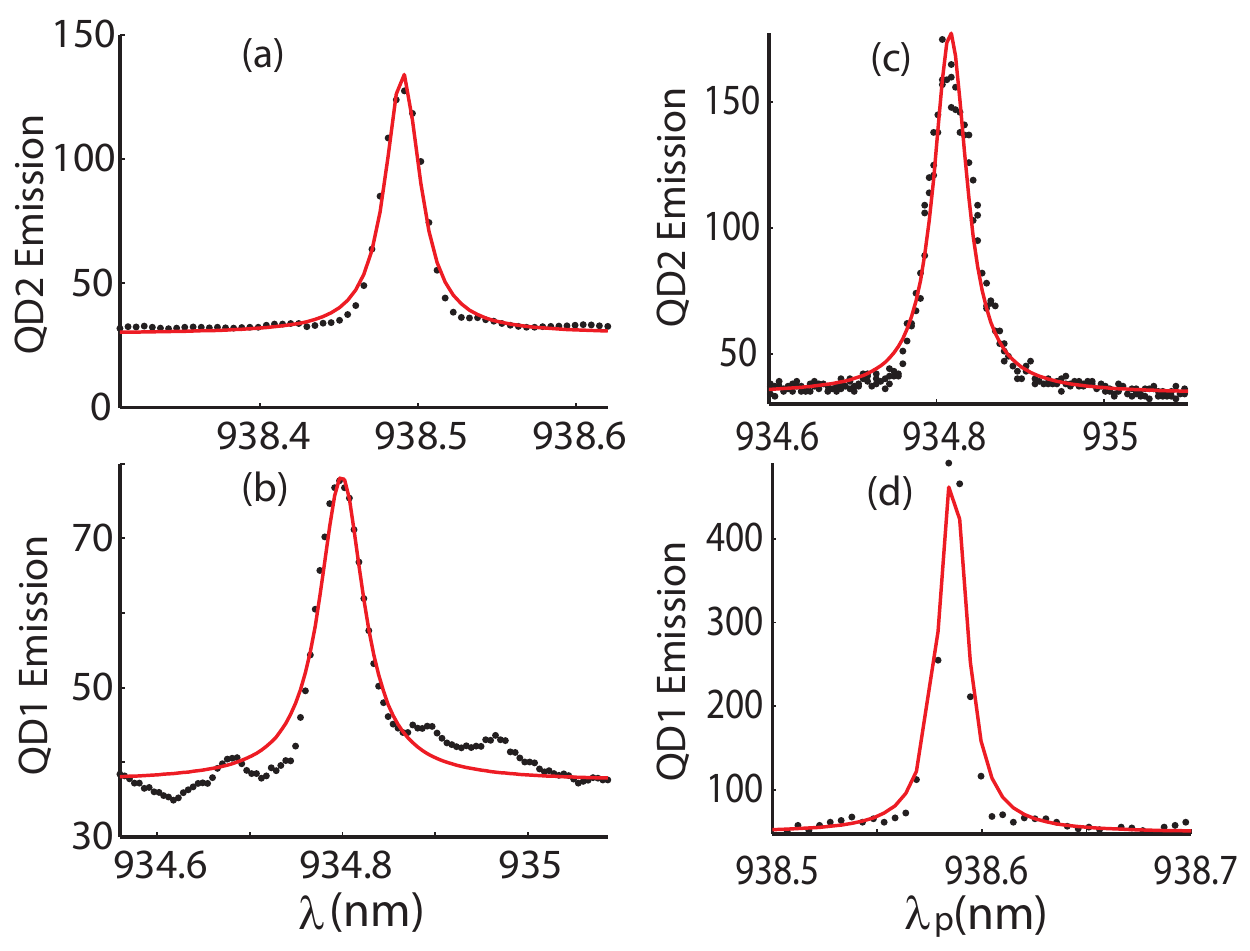}
\caption{(color online) Comparison of the linewidths measured in
direct off-resonant dot emission (Fig. \ref{Fig_exp_dot_dot_coupling_part_a}) and from resonant spectroscopy of
the QDs: (a),(b) the off-resonant QD emission (same as the
inset of Fig. \ref{Fig_exp_dot_dot_coupling_part_a}). QD
linewidths are estimated by fitting Lorentzians. Linewidths of the
higher and lower energy QDs, respectively, are $\sim 0.03$ nm and
$\sim 0.013$ nm. The cavity is at $\sim 935$ nm, close to the
higher energy QD. (c),(d) the off-resonant dot emission as a
function of the pump laser wavelength $\lambda_p$. In this experiment, a laser is scanned across one QD and emission is collected from the other QD as a function of the laser wavelength, same as in Ref. \cite{article:majumdar10}. By fitting
Lorentzians, we estimate the linewidths of the higher and lower
energy QDs to be $\sim 0.024$ nm and $\sim0.008$ nm,
respectively.} \label{Fig_exp_dot_dot_coupling_part_b}
\end{figure}

In the second system, the QD that is spectrally close to the
cavity is only weakly coupled to it. We observe emission from a
lower energy QD when a laser resonantly excites a higher energy QD
(Fig. \ref{Fig_exp_dot_dot_coupling_part_a} blue plot). We also
observe up-conversion, i.e., emission from the higher energy QD
under excitation of the lower energy QD (Fig.
\ref{Fig_exp_dot_dot_coupling_part_a} red plot). The energy
difference between the two QDs corresponds to a $~1.2$ THz
acoustic phonon. The cavity is at $\sim 935$ nm, closer to the
higher energy QD, although its emission is not distinctly
noticeable. The data is taken at $25$K. In the inset (replicated
in Fig. \ref{Fig_exp_dot_dot_coupling_part_b} a,b), we plot the
collected emission from a QD and estimate the linewidth of the other (excited) QD by a
Lorentzian fit. The higher and lower energy QDs, respectively,
have linewidths of $\sim 0.03$ nm and $\sim 0.013$ nm. These are
comparable to the linewidths of the self-assembled QDs
\cite{article:majumdar10}, and indicate that the coupling is
indeed between two QDs. The broader linewidth of the higher energy
QD is due to the presence of the cavity. Following this, we
perform a more accurate measurement of the linewidths of each QD by observing the peak amplitude
of the emission from the off-resonant dot as a function of the
pump laser wavelength $\lambda_p$ (Fig.
\ref{Fig_exp_dot_dot_coupling_part_b} c,d)
\cite{article:majumdar10}. From the Lorentzian fit, we estimate
that linewidths of the higher and lower energy QDs are,
respectively, $\sim 0.024$ nm and $\sim0.008$ nm. The slightly
smaller linewidths measured by the latter approach are due to the better spectral resolution offered by
this method \cite{article:majumdar10}.

\begin{figure}
\centering
\includegraphics[width=3.5in]{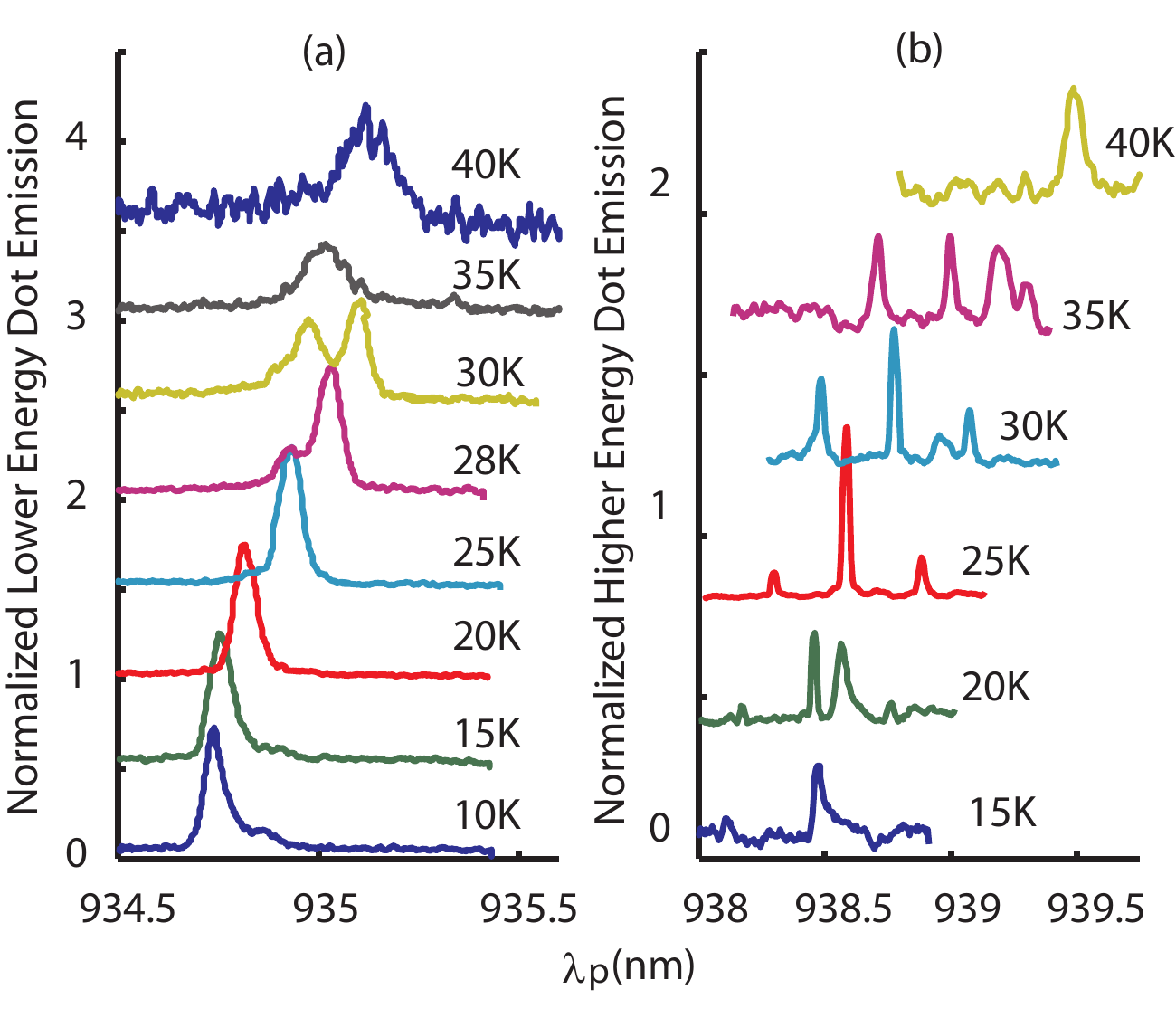}
\caption{(color online) Effects of temperature on dot to dot
coupling and the resulting frequency conversion of the pump. In
the experiments, the wavelength of the pump, $\lambda_p$, is scanned
through one QD and the peak intensity of the other QD is
monitored. In (a), the pump is scanned through the higher energy
QD and the down-converted light from the lower energy QD is collected and 
plotted, while in (b), we plot the up-converted light emitted from
the higher energy QD as the pump is scanned through the lower
energy QD. In both figures, the plots are vertically offset for
clarity. } \label{Fig_exp_dot_dot_temperature}
\end{figure}

Finally, we performed a study of the effects of temperature on the
inter-dot coupling. We note that while down-conversion of the pump
light is observed at a temperature as low as $10$ K,  we did not
observe any up-conversion at this temperature. This corroborates
the fact that the observed dot to dot coupling is phonon-mediated,
and at a low temperature up-conversion cannot happen due to the smaller
number of phonons. We first scan the laser across the higher
energy QD and observe the off-resonant emission from the lower
energy QD; Fig. \ref{Fig_exp_dot_dot_temperature}a shows the
result of this measurement for a set of different temperatures.
Similarly, Fig. \ref{Fig_exp_dot_dot_temperature}b shows the data
obtained by scanning the laser across the lower energy QD and
observing the off-resonant emission from the higher energy QD for
an assortment of temperatures. It can be seen from the
down-conversion plots in Fig. \ref{Fig_exp_dot_dot_temperature}a
that at lower temperature we observe emission from the lower
energy QD only when the pump is within the linewidth of the higher
energy QD. However, with increasing temperature we observe
emission from the lower energy QD even when the cavity is pumped.
When the temperature is raised to $\sim40$ K, we observe coupling
only from cavity to the lower energy QD, similar to the
observations reported previously \cite{article:majumdar09}. We
note that for the experimental result (performed at a temperature
of $\sim 45$ K) shown in Figure \ref{Fig_exp_sc_dot_temp} (for the
system only with a strongly coupled dot), we also observe coupling
between the cavity and the QD, and not between two QDs. This
disappearance of dot to dot down-conversion might be caused by the
increase in phonon density and the resulting broadening of the QD
lines. In Fig. \ref{Fig_exp_dot_dot_temperature}b, we monitor the
effects of temperature on the up-conversion. We cannot detect the
up-conversion at $10$ K, as it only becomes observable at higher
temperatures. However, with increasing temperature, the QD lines
disappear. This is most likely due to the fact that the QD starts
losing confinement with higher temperature. The additional peaks
in Fig.  \ref{Fig_exp_dot_dot_temperature}b show up-conversion of
several other QDs.

In summary, we observed phonon mediated inter-dot coupling, both
in systems with strongly and weakly coupled QDs. Both frequency
up- and down-conversion were reported via a phonon of estimated
energy $\sim 1.2$ THz. Our results indicate that this coupling is
enhanced by the presence of the cavity, and that without a cavity
spectrally close to one of the QDs this process does not occur.

\section{Acknowledgements}
The authors acknowledge financial support provided by the ONR, NSF,
and ARO. E.K. acknowledges support from the IC Postdoctoral
Research Fellowship. A.R. was supported by a Stanford Graduate
Fellowship. The authors acknowledge Dr. Pierre Petroff and Dr.
Hyochul Kim for providing the QD sample.
\bibliography{NRDC_bibl_PR}
\end{document}